\documentclass[9pt,journal]{IEEEtran}
\usepackage{ifpdf}
 \ifpdf
 \else
 \fi

\ifCLASSINFOpdf
   \usepackage[pdftex]{graphicx}
   \graphicspath{{../pdf/}{../jpeg/}}

   \DeclareGraphicsExtensions{.pdf,.jpeg,.png}
\else

\fi

\usepackage{amsmath}

\usepackage[style=ieee]{biblatex}
\usepackage[utf8]{inputenc} 
\usepackage{dblfloatfix}    
\usepackage[dvipsnames]{xcolor}
\newcommand{\squeezeup}{\vspace{-2mm}}

\begin{document}
\title{Step-Change in Friction under Electrovibration}

\author{Idil Ozdamar$^{1}$, M.Reza Alipour $^{1}$,  Benoit P. Delhaye$^{2,3}$, Philippe Lefèvre$^{2,3}$, and Cagatay Basdogan$^{1}$%
\thanks{$^{1}$ College of Engineering, Koc University, 34450, Istanbul, Turkey}%
\thanks{$^{2}$ Institute of Neuroscience, Université Catholique de Louvain, Brussels, Belgium}%
\thanks{$^{3}$ Institute of Information and Communication Technologies, Electronics and Applied Mathematics, Université Catholique de Louvain, Louvain-la-Neuve, Belgium}%
\thanks{$^{\#}$Corresponding author: \tt\small cbasdogan@ku.edu.tr}
}

\markboth{IEEE TRANSACTIONS ON HAPTICS}{}%

\maketitle

\begin{abstract}
Rendering tactile effects on a touch screen via electrovibration has many potential applications. However, our knowledge on tactile perception of change in friction and the underlying contact mechanics are both very limited. In this study, we investigate the tactile perception and the contact mechanics for a step change in friction under electrovibration during a relative sliding between a finger and the surface of a capacitive touchscreen. First, we conduct magnitude estimation experiments to investigate the role of normal force and sliding velocity on the perceived tactile intensity for a step increase and decrease in friction, called rising friction (RF) and falling friction (FF). To investigate the contact mechanics involved in RF and FF, we then measure the frictional force, the  apparent  contact  area, and the strains acting on the fingerpad during sliding at a constant velocity under three different normal loads using a custom-made experimental set-up. The results show that the participants perceived RF stronger than FF, and both the normal force and sliding velocity significantly influenced their perception. These results are supported by our mechanical measurements; the relative change in friction, the apparent contact area, and the strain in the sliding direction were all higher for RF than those for FF, especially for low normal forces. Taken together, our results suggest that different contact mechanics take place during RF and FF due to the viscoelastic behavior of fingerpad skin, and those differences influence our tactile perception of a step change in friction.
\end{abstract}



\IEEEpeerreviewmaketitle

\section{Introduction}

\IEEEPARstart{T}{he} technology for displaying tactile feedback through a capacitive touch screen via electrovibration is already in place and straightforward to implement, but our knowledge on the contact mechanics underlying electrovibration is still limited though some recent studies have shed some light on it \cite{Ayyildiz12668,adhession,Thonnard,Shultz}. Ayyildiz et al. \cite{Ayyildiz12668} investigated the sliding friction under electrovibration as a function of normal force and voltage using (i) a mean field theory based on multiscale contact mechanics, (ii) a full-scale computational contact mechanics study, and (iii) experiments performed on a custom-made tribometer. They showed that electroadhesion by electrovibration causes an increase in the real contact area at the microscopic level leading to an increase in the frictional force. They also argued that it is possible to further augment this force and, thus, the tactile sensation by using a thinner insulating film on the touchscreen. Sirin et al.\cite{Thonnard} conducted an experimental study to investigate the contact evolution under electrovibration using a robotic set-up and an imaging system originally developed by Delhaye et al. \cite{delhaye}. The results showed that the coefficient of friction increases under electrovibration as expected, but the apparent contact area is significantly smaller during sliding when compared to that of no electrovibration condition. Based on an adhesive friction model suggested in the literature \cite{Adams}, they also argued that the increase in friction force under electrovibration is actually due to an increase in the real contact area. Moreover, they showed that fingerpad moisture has an adverse effect on electrovibration. They speculated that high moisture decreased the electrical impedance of the interfacial gap between the finger and the touchscreen leading to a smaller increase in frictional force. Supporting this result, a recent experimental study by Shultz et al. \cite{Shultz} showed that the electrical impedance of the interfacial gap is significantly lower for the stationary finger compared to that of the sliding finger under electrovibration, suggesting that the role of moisture is reduced during sliding. These results on moisture, in fact, also explain why electovibration is not perceived well when a finger is stationary since the sweat accumulates in the gap and shorts out the applied voltage.  

In order to generate the desired tactile effects on a touch surface by friction modulation for displaying virtual shapes and textures, it is important to understand our perception of change in friction and the underlying contact mechanics. Numerous works have already examined the friction between the finger and surfaces of different roughness \cite{Derler2009,Kuilenburg,Sahli,delhaye2}, but the number of studies investigating  the  change  in  friction  on  the  same  surface are very  limited \cite{hurrem}. Changing friction via electrovibration is likely to generate local strains (compressive and tensile) in the fingerpad skin during sliding due to its viscoelastic nature. Human tactile afferents, which are the source of our tactile perception, have been shown to be exquisitely sensitive to local skin deformations \cite{Vallbo,Saal,Phillips} and are, therefore, likely to encode those events. However, the exact role that contact mechanics play in our tactile perception of change in friction under electrovibration have yet to be understood.

The objective of this study is twofold. First, we investigate our tactile perception for a step change in friction under electrovibration. We then measure the underlying contact mechanics and relate them to the perception. To do so, we measure tangential friction force, apparent finger contact area, and the strains acting on the fingerpad under RF and FF conditions for 3 different normal loads while a capacitive touch screen slides at a constant velocity under the fingerpad. The results of our experiments with 10 participants showed that rising friction (RF) is perceived stronger than falling friction (FF). Accordingly, our contact mechanics analysis, performed with another group of 10 participants, revealed that the contrasts in coefficient of friction, apparent contact area, and mechanical strains before and after the friction change are all higher for RF than those for FF, and the viscoelastic behavior of the fingerpad skin is potentially responsible for the difference in our tactile perception.

\section{Material and Methods}
\subsection{Experimental Apparatus}
Our experimental set-up (Figure 1), inspired by Tada et al. \cite{Tada} and Delhaye et al. \cite{delhaye}, was developed to investigate friction between a fingerpad and a touch screen and distortion of the fingerpad as the finger slips. This system consists of a capacitive touch screen (SCT3250, 3M Inc.), a co-axial light source (C50C, Contrastech Inc.), a high-speed camera (IL5H, Fastec Imaging Inc.) and a force transducer (Mini40-SI-80-4, ATI Inc.) placed under the touch screen, and two linear translational stages (LTS150, Thorlabs Inc.) that enable to control the movements of the touch screen along Z (using stage 1) and Y (using stage 2) axes as shown in Fig.1. The acceleration of both stages was 50 mm/s$^2$. Tangential and normal force acting on the fingerpad were measured by the force sensor and sampled by a data acquisition card (PCIe-6034E, National Instruments Inc.) at a frequency of 10 kHz. A proportional-integral-derivative (PID) controller was implemented to keep the normal force acting on the fingerpad at a constant value. The PID loop was updated at a rate of 50 Hz. The touch screen was excited with a sinusoidal voltage signal generated by a data acquisition card (PCIe-6321, National Instruments Inc.) and augmented by a piezo amplifier (E-413, Physik Instrumente Inc.). A custom-made hand support was manufactured using 3D printing techniques to keep the participant index finger in place and to provide a comfortable hand position. The contact angle relative to the surface was set to $20^o$. A computer fan continuously blew air over the touch surface to minimize the undesirable effects of moisture. The contact area was imaged by the high-speed camera through the coaxial light source. Finger images were captured by the camera at a speed of 100 fps (frames per second) and resolution of 1920 x 1080 pixels. The images were stored in the memory after each slip using the auto-save feature of the camera. The coaxial light module contains a LED array, a half mirror, and a diffuser to acquire finger images in high contrast. The cold white light beam emitted from the LED array was reflected by the half mirror inside the light source, reached to the touchscreen, absorbed at the ridges and reflected at the valleys of the fingerprint, and then captured by the camera. Thus, the fingerpad ridges appeared darker than valleys in the captured images.   
\begin{figure}[!t]\squeezeup
\centering
\includegraphics[width=1\linewidth]{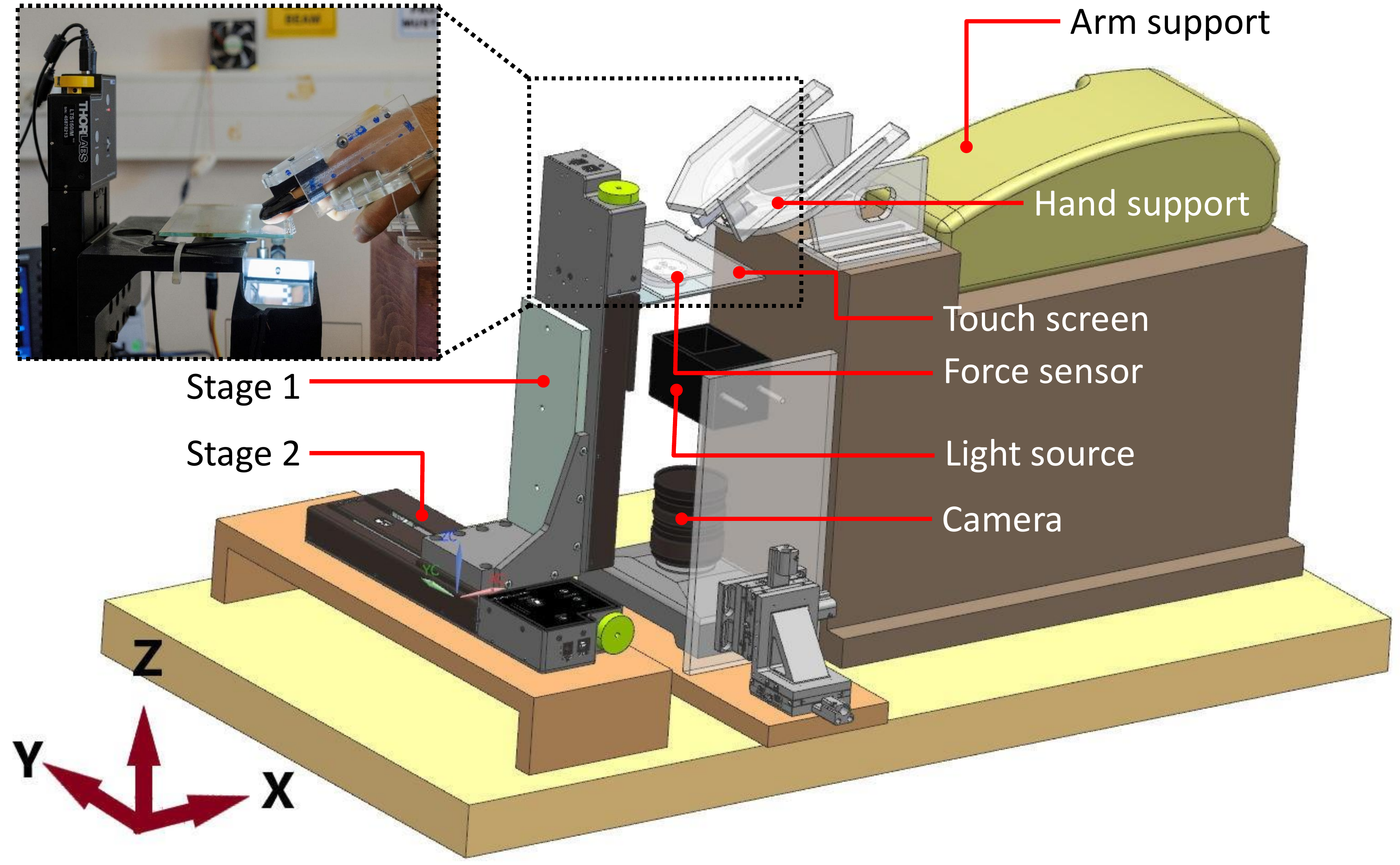}\squeezeup
\caption{Experimental set-up. The participant's index finger is secured in a hand support and stimulated by electrovibration displayed through a flat transparent touch screen. The normal force and sliding velocity are controlled by stage 1 and 2 respectively. The contact area is imaged by a high-speed camera using a coaxial light source.}\squeezeup
\label{fig_sim}
\end{figure}\squeezeup

\subsection{Participants}
The perception and contact mechanics experiments were conducted seperately with two different groups of 10 participants with dry fingers. The average age of the participants in the perception experiment and contact mechanics experiment were $26.5\pm{7.6}$ and $25.3\pm{4.4}$ years, respectively. To avoid the undesirable effects of moisture, the participants of both experiments were selected from a large pool of potential candidates based on their fingertip moisture levels. For this purpose, the fingertip moisture of each candidate was measured by a Digital Skin Analyzer (SK-8, V-Care Inc.) 3 times in 3 consecutive days before the experiments. Based on these measurements, a moisture level of 60 (a. u.) was used as a threshold to group the candidates. The average moisture levels of dry and moist candidates were $51.1\pm{4.4}$ and $88.5\pm{2.1}$, respectively. Only the candidates in the dry group were selected for the experiments. Before starting the experiments, the touch screen was cleaned with alcohol and the participants washed their hands with commercial soap, rinsed with water, and dried them at room temperature. To ensure a good contrast in the captured images, 0.5 $\mu$L of oil (liquid Vaseline) was carefully applied to the participants’ fingertips using a micro pipette. While oil reduced the coefficient of friction at the interface, recording finger images in high contrast was not possible without it since the touch screen used in this study was less transparent than the standard glass surfaces used in earlier studies \cite{delhaye}. In order to reduce the potential adverse effects of oil on our measurements, we used metrics emphasizing the contrast between RF and FF rather than the absolute values. All participants read and signed a consent form before the experiment. The form was approved by the Ethical Committee for Human Participants of Koc University. All procedures performed in this study were done in accordance with the Declaration of Helsinky and the experiment was performed in accordance with relevant guidelines and regulations.

\begin{figure*}[!t]\squeezeup\squeezeup
\includegraphics[width=1\textwidth]{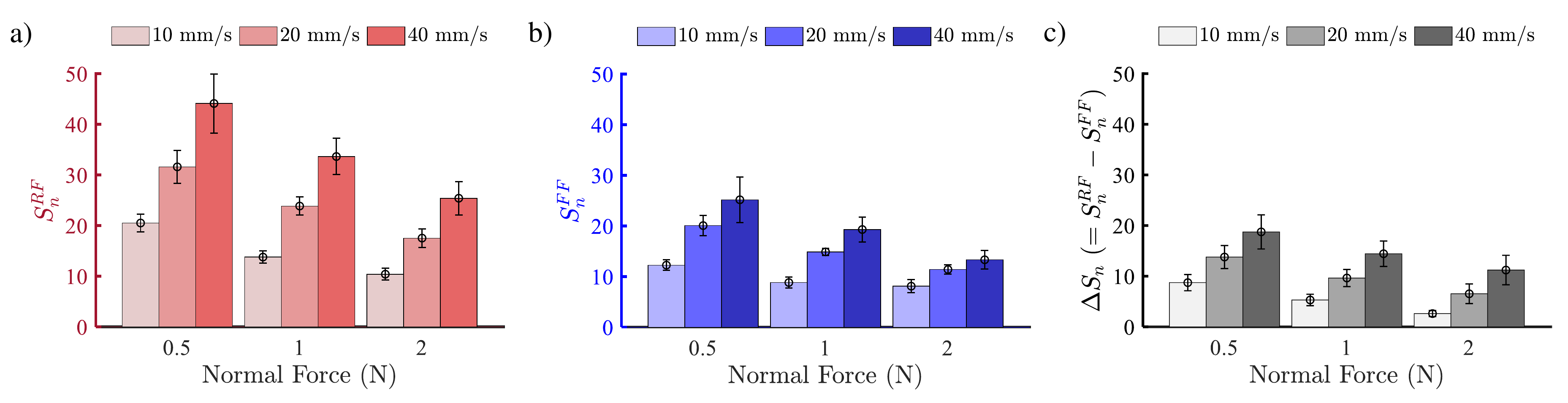}\squeezeup\squeezeup
\caption{Results of the perception experiment. Mean (average of 10 participants) of normalized intensity scores ($S_{n}$) for RF (a) and FF (b), and the difference between RF and FF (c) for 3 different normal forces (0.5, 1.0 and 2.0 N) and 3 different sliding velocities (10, 20 and 40 mm/s). The error bars show the standard error of the means.}\squeezeup
\end{figure*}
\begin{figure*}[!b]
\centering
\includegraphics[width=1\linewidth]{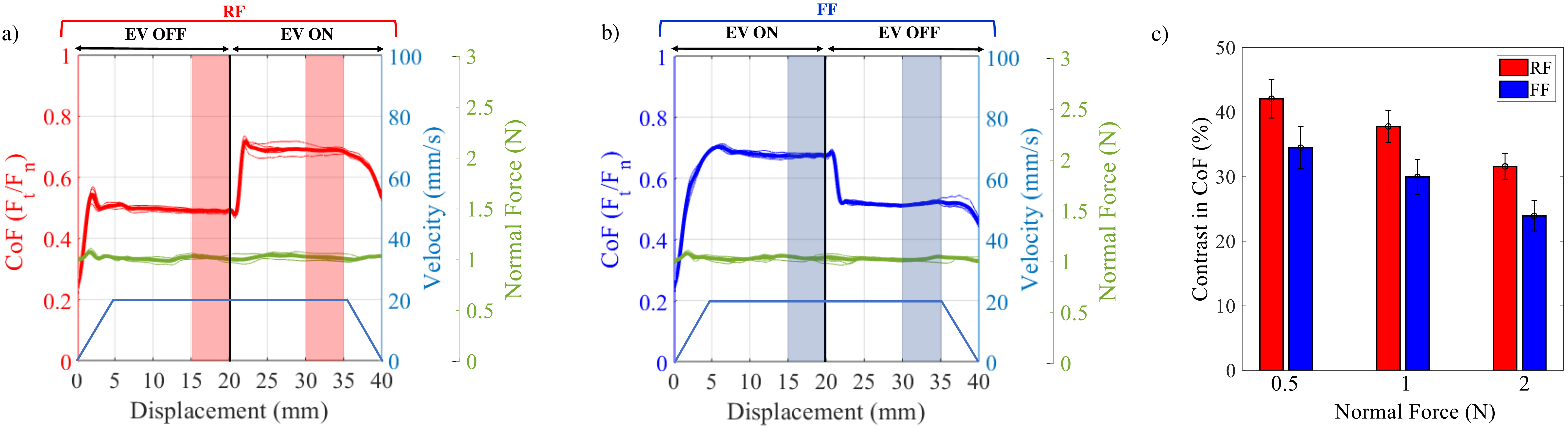}\squeezeup
\caption{Change in coefficient of friction (CoF) of one participant as a function of displacement under RF (a) and FF (b) conditions for the normal force of 1.0 N. The y-axis on the right in green (blue) color is for the recorded normal force (velocity profile). The mean (average of 10 participants) normal force applied by the participants (regulated by a PID control) were $0.51 \pm 0.03$, $1.02 \pm 0.05$ and $2.03 \pm 0.07$ N. The abbreviation EV in plots stands for electrovibration. Thin lines show the individual slips and thick lines are the average of 5 slips. The shaded regions in light red and light blue show the displacement intervals where the CoF reaches steady-state value. (c) Mean (average of 10 participants) contrast in CoF under RF and FF for 3 different normal forces. The error bars show the standard error of means.}\squeezeup
\centering
\label{friction_plots}
\end{figure*}
\subsection{Experimental Procedure}
The participants placed their index finger into the hand support (Fig. 1) to make a constant angle of contact ($20^o$) with the touchscreen. First, stage 1 was moved up slowly along the vertical direction (Z-axis) until the touch screen made contact with their fingerpad and the normal force acquired by the force sensor reached the desired value. Then, stage 2 was commanded to move the touch screen under the fingerpad along the tangential direction (Y-axis) for a travel distance of 40 mm at a constant velocity while the normal force was kept constant at the desired value via the PID controller. For RF, a sinusoidal voltage signal with an amplitude of 300 $V_{pp}$ and frequency of 125 Hz was applied to the screen right after crossing the mid-point of the travel distance. For FF, the same voltage signal was applied to the screen from the beginning of the travel until the mid-point was reached and was then turned off.  

\underline{Perception experiment:} The experiment was repeated for 3 different normal forces (0.5, 1.0 and 2.0 N) and 3 different velocities (10, 20 and 40 mm/s). Hence, each participant performed 108 trials in 3 sessions: 3 normal forces x 3 velocities x 2 experimental conditions (RF vs FF) x 6 repetitions. The order of the trials was randomized while the same order was displayed to each participant. There was a one minute break after each session (i.e. every 36 trials). The participants were asked to wear headphones during the experiment, and a white noise was played through the headphones to prevent any perceptual bias due to an external auditory noise. After each trial, the participants were asked to rate the tactile intensity of friction change. They were asked to report their scores using a small numpad by entering any positive number. The intensity scores of each participant were normalized using the method suggested by Murray et al. \cite{Murray}. For this purpose, the geometric mean of raw intensity scores of each participant, $GM_{participant}$, and the geometric mean of all participants, $GM_{all}$, were calculated. Finally, the normalized scores of each participant ($S_{n}$) were calculated by multiplying her/his raw scores with the ratio of $GM_{all}/GM_{participant}$.

\underline{Contact mechanics experiment:} The experiment was repeated for 3 different normal forces (0.5, 1.0 and 2.0 N) at a constant sliding velocity of 20 mm/s. Hence, each participant performed 30 slips: 3 normal forces (0.5, 1.0 and 2.0 N) x 2 experimental conditions (RF vs FF) x 5 repetitions. The experimental procedure was the same as the perception experiment. The tangential force acting on each participant's finger was recorded as the touch screen was moved by stage 2. The software command that activated the movement of stage 2 also triggered the camera to start recording fingerprint images. After each slip, the images were automatically saved to SSD (Solid State Disk) and the camera was armed to trigger again for the next slip.
\subsection{Metrics Used for Contact Mechanics Analysis}
\underline{Coefficient of friction:} The coefficient of friction (CoF) was calculated using the relation $CoF = F_{t}/F_{n}$, where $F_{t}$ is tangential force and $F_{n}$ is normal force.  

\underline{Apparent Contact Area:} The apparent contact area was calculated from the captured images using the approach suggested by Delhaye et al. \cite{delhaye2}. This approach involves a) applying a filter to the raw image data, b) identifying fingerpad contours, and c) fitting an ellipse to the contours. The area of the fitted ellipse was taken as the apparent contact area.

\underline{Strains:} The distribution of surface strain over the contact area was also calculated using the approach suggested by Delhaye et al. \cite{delhaye2}. First, feature points were sampled from the contact area, and then Delaunay triangulation was implemented to connect the feature points and form a triangular mesh. Second, the instantaneous strain values (normal: $\epsilon _{xx}$, $\epsilon _{yy}$, and shear: $\epsilon _{xy}$) were calculated by tracking the deformation of each triangle in consecutive image frames, and then assigned to the center of triangles. The accumulated strain between any two non-consecutive frames for each triangle was obtained by integrating the instantaneous strains. Strain distribution over the contact area was obtained by calculating the accumulated strains for all triangles. As a result, finger deformation in the contact area was determined for each strain component. In order to report the average strain values across all participants, a normalization was applied to the strain values of each participant due to the differences in their contact areas and, hence, the number of triangles being tracked. For this purpose, the strain values assigned to the triangles of each participant were mapped to a rectangular grid with constant dimensions by interpolation.
\squeezeup
\section{Results}
First, we compared the perceived intensity for a step change in friction under RF and FF conditions. Then, we analyzed the relative changes (contrast) in the coefficient of friction and the apparent contact area due to RF or FF. Finally, we reported the surface strains along the direction of sliding ($\epsilon _{xx}$). 
\subsection{Perceived intensity for a step change in friction}
The normalized intensity scores of the participants for RF ($S_{n}^{RF}$) and FF ($S_{n}^{FF}$) are reported in Figs. 2a and 2b, respectively. The scores were analyzed using a three-way analysis of variance (ANOVA) with repeated measures using a) experimental condition (RF, FF), b) normal force ($F_{n} = 0.5, 1.0, 2.0$ N), and c) velocity ($v = 10, 20, 40$ mm/s) as the main factors. The analysis showed that RF was perceived significantly stronger than FF ($F_{1,9} = 19.15$, $p = 0.002$). As shown in Fig. 2c, the difference between intensity scores ($\Delta S_{n}=S_{n}^{RF}-S_{n}^{FF}$) was always positive. Sliding velocity and normal force significantly affected the participants' intensity scores ($S_{n}$). An increase in sliding velocity increased intensity score ($F_{2,18} = 14.10$, $p = 0.003$) while a decrease in normal force also increased the score ($F_{2,18}= 20.84$ , $p = 0.001$). There was a statistically significant interaction between the experimental condition and normal force ($F_{2,18} = 8.37$, $p = 0.007$) and also between the experimental condition and velocity ($F_{2,18} = 12.67$, $p = 0.001$), which was further analyzed by Bonferroni-corrected paired t-tests and found significant for all pairs ($p < 0.05$). Finally, no significant interaction was observed between velocity and normal force. 

\begin{figure}[!b]
\centering
\includegraphics[width=1\linewidth]{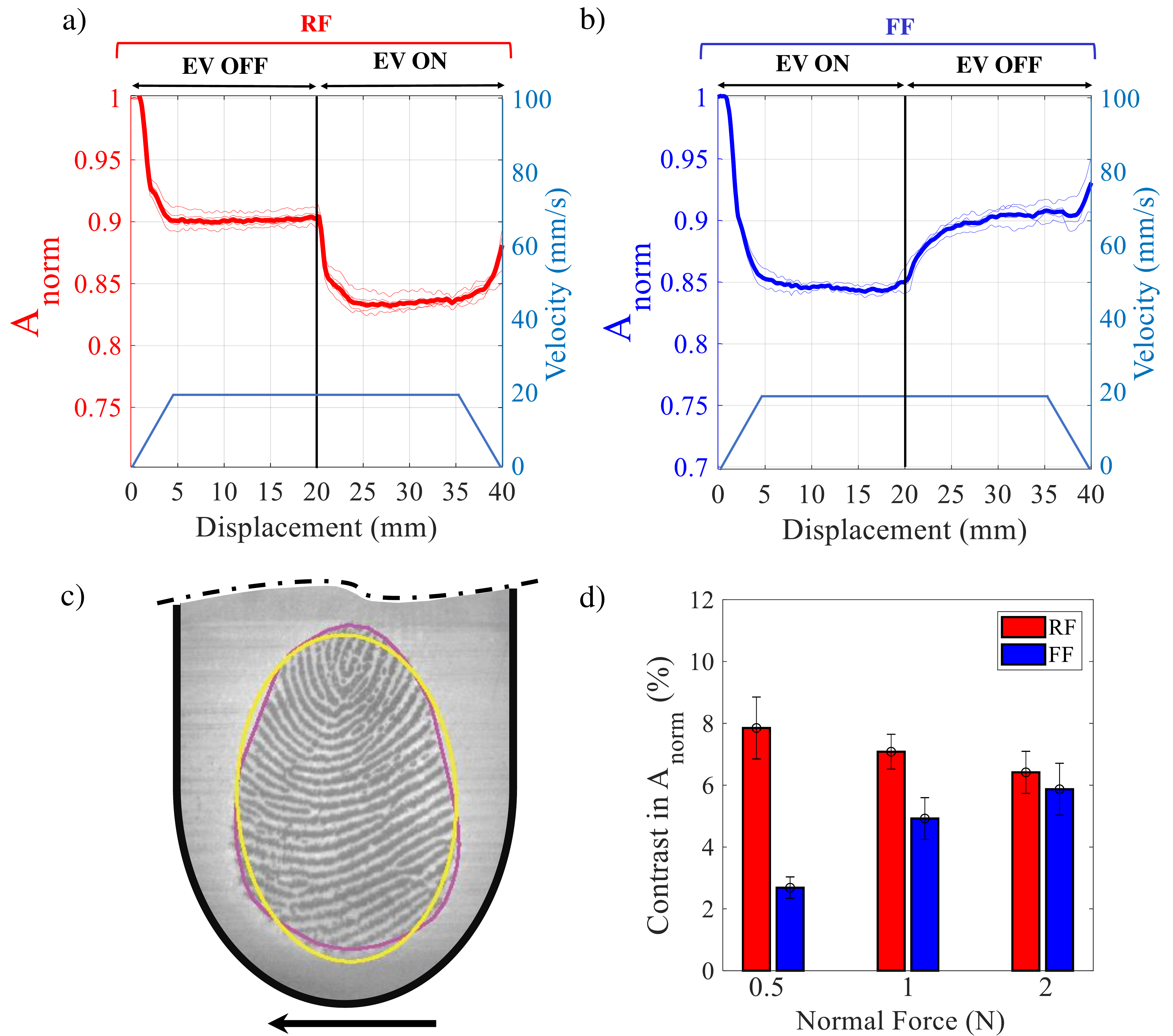}
\squeezeup
\caption{Change in apparent contact area during RF and FF. Normalized apparent contact area of one participant as a function of relative displacement under RF (a) and FF (b) for a normal force of 1.0 N. 'EV' stands for electrovibration, thin lines represent the individual slips and thick lines represent the average of 5 slips. The y-axis on the right in blue color is for velocity profile. (c) Captured image of a fingerpad in contact with the touchscreen; the contour around the contact area is in purple while the ellipse fitted to the contour is in yellow. The arrow under the finger image shows the direction of touch screen movement (radial direction). (d) Mean (average of 10 participants) contrast in normalized apparent contact area under RF and FF for 3 normal forces. The error bars show the standard error of means.}\squeezeup
\centering
\label{area_plots}
\end{figure}

\begin{figure*}[!t]
\centering
\includegraphics[width=1\linewidth]{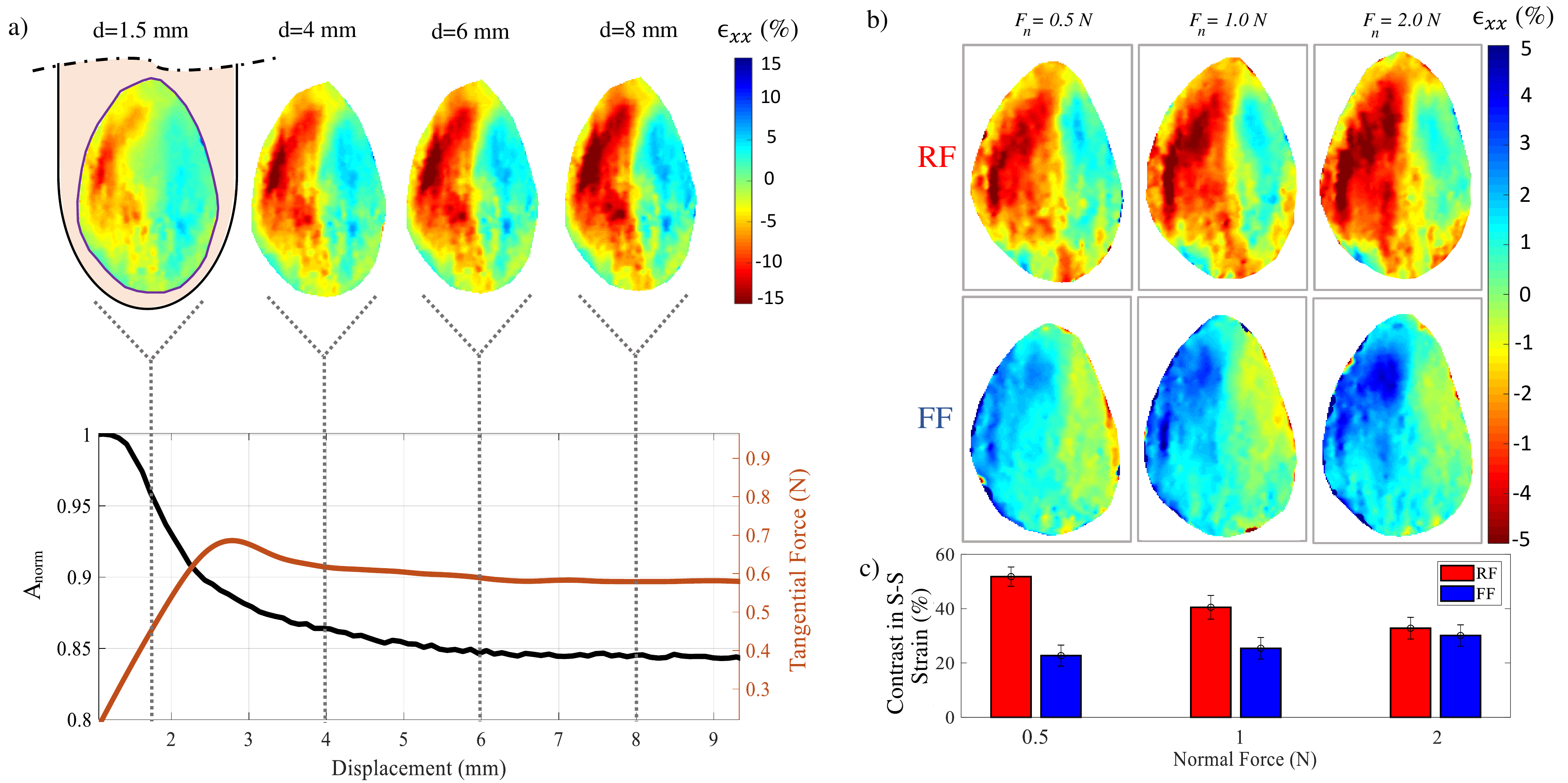}\squeezeup
\caption{Strain distribution and mean strains along the direction of finger motion ($\epsilon _{xx}$) (a) Top: Evolution of strain distribution as a function of displacement for one participant covering the range from the initiation of motion to the early stages of sliding when EV is OFF. Bottom: Evolution of the contact area and the tangential force as a function of displacement for the same example trial. (b) Difference between the steady-state strains acting on the fingerpad of the same participant before and after the friction change under RF (top row) and FF (bottom row) for three different levels of normal force. (c) Mean (average of 10 participants) contrast in steady-state (s-s) strains (integrated over the contact area) under RF and FF. The error bars show the standard error of mean.}\squeezeup
\centering
\label{area_plots1}
\end{figure*}
\begin{figure}[!b]
\centering
\includegraphics[width=1\linewidth]{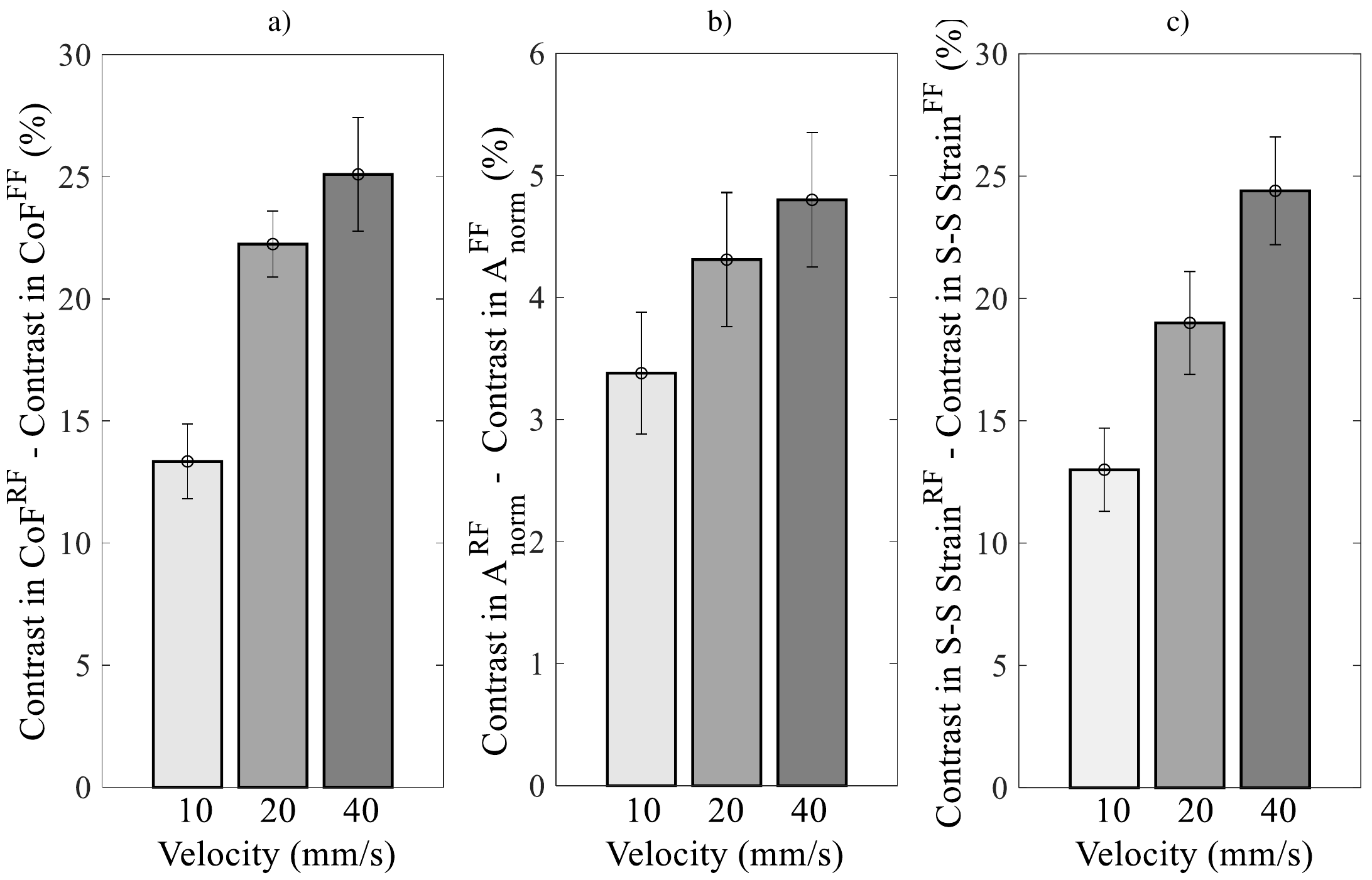}\squeezeup

\caption{Difference between the contrast metrics calculated for RF and FF of one participant as a function of sliding velocity for the normal force of 1.0 N. Difference between the contrasts in (a) CoF, (b) normalized contact area and (c) steady-state (s-s) strain. The error bars represent the standard deviations from the mean values}\squeezeup
\centering
\label{energy}
\end{figure}
\subsection{Contact mechanics for a step change in friction}
\underline{Coefficient of friction:} Figure 3 shows the change in CoF of one participant as a function of displacement for a normal force of 1.0 N under RF (Fig. 3a) and FF (Fig. 3b). The shaded regions in the plots are the displacement intervals where CoF reaches the steady-state value. The contrast in CoF, ($1-CoF_{OFF}/CoF_{ON}$) x 100, was calculated by using the average values of CoF in those regions. Fig. 3c shows the average contrast values for all participants. A two-way ANOVA with repeated measures was performed on the friction contrast using a) the experimental condition (RF, FF) and b) normal force ($F_{n} = 0.5, 1.0, 2.0 $ N) as the main factors. Both the experimental condition ($F_{1,9} = 51.74$, $p < 0.001 $) and normal force ($F_{2,18} = 32.41$, $p < 0.001$) significantly affected the contrast in CoF. There was no statistically significant interaction between the main factors.

\underline{Apparent Contact Area:} 
Figure 4 reports the normalized apparent contact area ($A_{norm}$ = $A_{app}/A_{0}$, where $A_{0}$ is the initial apparent contact area) of one participant as a function of finger displacement under RF (Fig. 4a) and FF (Fig. 4b) for a normal force of 1.0 N. The normalization helps to evaluate the change in apparent contact area across the participants. We observed that the apparent contact area of each participant reached a steady-state value at the shaded regions (Figs. 3a and 3b) under both experimental conditions (RF and FF). Again, the contrast in contact area, ($\mid$$1-A_{norm,OFF}/A_{norm,ON}$$\mid$) x 100, was calculated using the average values of apparent contact areas in those regions. Fig. 4d shows the average contrast values for all participants. A two-way ANOVA with repeated measures was performed on the contrast in apparent contact area using a) the experimental condition (RF, FF) and b) normal force ($F_{n} = 0.5, 1.0, 2.0$ N) as the main factors. The experimental condition ($F_{1,9} = 55.98$, $p < 0.001$) and normal force ($F_{2,18} = 4.49$, $p = 0.04$) significantly affected the contrast in apparent contact area. Moreover, there was a statistically significant interaction between the experimental condition and normal force ($F_{2,18} = 31.44$, $p < 0.001$), which was further analysed by Bonferroni-corrected paired t-tests and found significant for the normal force values of 0.5 N and 1.0 N ($p < 0.001$).

\underline{Strains:}
First, we investigated the evolution of compressive and tensile strains (along the axis of motion, $\epsilon_{xx}$) at the onset of the movement when EV is OFF. The strain distribution on the fingerpad of one participant is depicted in Fig. \ref{area_plots1}a for 4 successive frames, sampled from the frames covering the interval ranging from initiation of motion to the steady-state sliding. As reported previously \cite{delhaye2}, compressive (red colored) and tensile (blue-colored) strains were accumulated on the radial and ulnar portions of the fingerpad (respectively) as the touch screen was moved in the radial (towards thumb) direction. Next, we investigated the difference in strain distribution resulting from RF (see Fig. \ref{area_plots1}b, top row) and FF (Fig. \ref{area_plots1}b, bottom row). In order to construct these maps, the strain distribution corresponding to the steady-state region before the friction change was subtracted from the one after the friction change. We observed that RF and FF had opposing effects on fingerpad deformations. Indeed, RF caused an "overstress". That is, the regions of fingerpad that had been compressed at the onset of movement were further compressed. On the other hand, FF resulted in stress relaxation. Indeed, the regions in compression at the onset of movement were partially relaxed and, hence, the difference in steady-state strains before and after the friction change in those regions was positive (tensile) for FF. Furthermore, the amplitude of the strains increased with normal force in both cases. The mean (averaged over 10 participants) contrast in strain, integrated over the contact area, under RF and FF is reported in Fig. \ref{area_plots1}c. The contrasts were computed using the steady-state strain values before and after the friction change. Using a two-way ANOVA with repeated measures, we observed that the experimental condition ($F_{1,9} = 83.07$, $p < 0.001$) and normal force ($F_{2,18} = 8.55$, $p = 0.002$) had a significant effect on the contrast in strain. Moreover, there was a statistically significant interaction between the experimental condition and normal force ($F_{2,18} = 71.75$, $p < 0.001$), which was further analysed by Bonferroni-corrected paired t-tests and found to be significant for the normal force values of 0.5 N and 1.0 N ($p < 0.05$).

\squeezeup
\section{Discussion}
Our study showed that humans perceive RF stronger than FF, and the perceptual difference between these conditions can be explained by the difference in their contact mechanics. We argue that this difference stems from the viscoelastic nature of human fingerpad skin. The experimental study conducted by Friesen et al. \cite{viscoFingerColgate} using different artificial fingertips supports this claim. Indeed, they compared the friction reduction ability of an artificial viscoelastic fingertip with the one that does not exhibit friction reduction, and they concluded that mechanical damping characteristics play a key role in the amount of friction reduction achieved.

In particular, our perception experiment showed that RF was perceived stronger than FF (Fig. 2). Moreover, the intensity scores of the participants increased with sliding velocity under both conditions, which is a sign of viscoelastic effects. The contact mechanics measurements supported the results of our perception experiment. We observed that the contrast in CoF was significantly higher under RF than that of FF (Fig. 3c). In addition, the contrast in CoF for both RF and FF conditions showed a decreasing trend as the normal force was increased, as reported in the literature earlier for finger-glass interface without EV \cite{Derler2009} and more recently with EV \cite{Thonnard}. This is also in-line with the trend observed in the intensity estimations of participants as a function of normal force (Fig. 2). The scores were inversely correlated with normal force for both RF and FF conditions. These results clearly show that the effect of electrovibration is more prominent at lower normal forces. 

Another metric used to compare RF and FF was the apparent contact area, $A_{app}$, which was estimated from the fingerprint images of participants captured by a high speed and resolution camera. First of all, our results showed a decrease in the apparent contact area at the initiation of the movement even when there was no EV probably due to the stiffening of finger skin in tangential direction  as reported in earlier studies \cite{delhaye}. In the case of RF, we observed a further decrease in contact area due to the increase in tangential force after EV was turned ON at the mid-point of traveling distance. The additional decrease in apparent contact area due to EV has already been reported by Sirin et al. \cite{Thonnard}. They speculated that the main cause of the increase in friction due to EV was an increase in real contact area and the reduction in apparent area is due to further stiffening of the finger skin in tangential direction. Symmetrically, in FF condition, the apparent contact area increased after EV was turned OFF at the mid-point of travel since the tangential force decreased and the fingerpad tissue relaxed. In addition, our current study showed that the contrast in apparent contact area under RF was higher than that of FF (Fig. 4d) while the gap between them reduced as the normal force was increased. This reduction in the gap at higher forces is not surprising since our study also shows that electroadhesion is more effective at lower normal forces.

Finally, we investigated the contact interactions between fingerpad and touch screen in detail using the strain distribution on fingerpad over the contact area. For all three normal forces, the contrast in steady strain was higher under RF compared to FF (Fig. 5c). The difference between the contrasts of RF and FF was again higher for lower normal forces. Fig. 5b shows the difference between the steady-state strain distributions before and after the friction change under RF and FF. A careful inspection of these maps in Fig. 5b in tandem with the evolution of strain maps in Fig. 5a for the initial stages of finger movement reveal that further compressive loading was applied to the fingerpad under RF due to the increase in tangential force. However, in the case of FF (Fig. 5b), the fingerpad skin started to relax since the initial compressive loading was reduced due to the decrease in tangential force and hence the difference between the steady-state strains before and after the friction change was positive (see the blue-colored tensile strain regions in the strain maps of FF in Fig. 5b). This relaxation behavior is typical of viscoelastic materials such as the fingerpad skin in our case. To further support our hypothesis on viscoelasticity, we present the contrast results of one participant for 3 different sliding velocities and the normal force of 1.0 N in Fig. 6. As shown in this figure, the difference in contrasts between RF and FF increases with the sliding velocity for CoF, apparent contact area, and steady-state strain.    
\squeezeup
\section{Conclusion}
Tactile perception of electrovibration and the underlying contact mechanics are currently under investigation. Understanding how we perceive a change in friction under electrovibration and the contact parameters affecting our perception can help user interface designers to display more realistic tactile feedback through touch surfaces. In this study, the difference between rising and falling friction (RF and FF), induced by electrovibration, was investigated for the first time. Under RF and FF, perceived tactile intensity, tangential friction force, apparent finger contact area, and the fingerpad surface strains were measured. Based on the data collected from 10 participants, we concluded that the contact mechanics of RF is significantly different than that of FF, and the difference is likely due to the viscoelastic nature of fingerpad skin. It is important to emphasize here that we allowed sufficient time interval to observe steady-state behaviour in CoF before and after the step change in friction for making comparison between RF and FF. The dynamics of multiple step changes in friction is different from the single one since the interval between the steps is a factor affecting the tactile perception. This topic (multiple step changes in friction) has been already investigated by Vardar et al. \cite{yaseminRoughness} for electrovibration and Gueorguiev et al. \cite{Gueorguiev2019} Saleem et al \cite{hurremToAppear} for ultrasonic actuation.

Here, one may argue that electrical effects also play a role in the difference. That is, some residual charges might have been left on the touch screen under FF when the voltage was turned OFF after the mid-point of travel distance, making electroadhesion still active for awhile in the OFF region. However, if the electrical model introduced in \cite{electricalModel} is considered, one can observe that the time constant for discharging after the EV is turned OFF is very small (compared to the mechanical relaxation time constant of fingerpad skin) and, hence, it is unlikely that there was a significant amount of residual charge left on the screen under FF after the EV was turned OFF. Moreover, Saleem et al. \cite{hurrem} conducted psychophysical experiments with an ultrasonic tactile surface display (which modulates the friction based on the principle of squeeze film effect) and found that RF was perceived stronger than FF. They also looked into the correlations between the perceived intensities of participants and several parameters involved in contact, and showed that the contrast and rate of change in tangential force were best correlated with the perceived intensity. Their results and the ones obtained in this study clearly demonstrate the important role that contact mechanics play in our tactile perception. 

In the future, we aim to develop mechanics-based fingerpad models and to investigate its contact interactions with a touch screen under electroadhesion to further support our claim on viscoelasticity and, also, to investigate the mechanical and electrical parameters affecting the contact interactions and our tactile perception in more depth.
\squeezeup
\section*{Acknowledgment}
We acknowledge the financial support provided by the Scientific and Technological Research Council of Turkey (TUBITAK) under contract number 117E954. We also acknowledge M. Latif Satir for his contributions to the design of the experimental setup.
\squeezeup
\renewcommand*{\bibfont}{\small}
\printbibliography

\end{document}